\documentclass[EPJ]{webofc}
\usepackage[varg]{txfonts}   
\wocname{EPJ Web of Conferences}
\woctitle{MENU 13}
\begin{document}
\title{An overview of neV probes of PeV scale physics}
\subtitle{-- and of what's in between}

\author{Susan Gardner\inst{1}\fnsep\thanks{\email{gardner@pa.uky.edu}} \and
        Brad Plaster\inst{1}\fnsep\thanks{\email{plaster@pa.uky.edu}} 
}

\institute{Department of Physics and Astronomy,
University of Kentucky, Lexington, KY 40506-0055 USA}

\abstract{%
Low-energy experiments which would identify departures
from the Standard Model (SM) rely either on the 
unexpected observation of symmetry breaking, 
such as of CP or B, or on an observed significant
deviation from a precise SM prediction. 
We discuss examples
of each search strategy, and show that low-energy experiments 
can open windows on physics far beyond accessible collider energies. 
We consider how the use of a frequentist analysis framework can 
redress the impact of theoretical uncertainties 
in such searches --- and how lattice QCD can help control them. 
}
\maketitle
\section{Context}
\label{preamble}
Direct searches for new physics at the LHC 
has yielded the discovery of the Higgs 
boson~\cite{Aad:2012tfa,Chatrchyan:2012ufa}, but no
unanticipated, new particles --- as yet. 
On the other hand, observational 
cosmology, analyzed in the framework of general relativity, 
tells us that only 4\% of the energy density
of the universe is in the matter we know~\cite{Ade:2013qta}, so that the SM 
of particle physics, successful though it is, is 
probably incomplete. The lack of evidence thus far for new
physics and interactions through collider studies at the highest
energies motivates broader thinking in the search for new physics. 
For example, the missing matter could be 
weakly coupled, making it more challenging if not impossible 
to identify in a collider environment. 
Low-energy, precision searches for new physics 
can also probe this alternative possibility 
and thus play 
a key role in the search for new physics. 
In this contribution we offer a terse overview of the
diverse program these experiments comprise. 

Generally, there are two distinct search strategies. 
That is, one can either make null tests of the breaking of
SM symmetries, or refine the measurement of 
quantities which can 
be computed, or assessed, precisely with the SM. 
In the former case, one can test, e.g., 
B-L invariance by searching for $n{\bar n}$ oscillations
or neutrinoless double-$\beta$ decay. Although CP 
is not a symmetry of the SM, there are nevertheless
observables for which the SM prediction is so small
that searches at current levels of sensitivity also constitute
null tests. Searches for permanent 
electric dipole moments (EDMs) of the neutron or electron, e.g., 
or for CP violation in 
the charm sector, be it through $D{\bar D}$ mixing or 
decay rate asymmetries, are examples of such tests. 
There are also a variety of nonzero observables whose value can be tested 
precisely within the SM. 
Example of this latter class include 
(i) parity-violating electron scattering 
from electrons, protons, or light nuclei, in varying kinematics, 
(ii) the anomalous magnetic moment of the muon (or electron), 
and (iii) final-state angular correlations in neutron and nuclear
$\beta$ decay. 
All these studies probe the possibility of new degrees of freedom, 
including those which couple so weakly to known matter 
that they are effectively
``hidden''. 

\section{Motivation}
\label{motive}
The SM leaves many questions unanswered; e.g., it 
cannot explain the nature of dark matter or dark energy, nor can it
explain the magnitude, or even existence, 
of the cosmic 
baryon asymmetry 
(BAU). 
The BAU itself can be determined by 
confronting the 
observed $^2$H abundance with big-bang nucleosynthesis, yielding 
$\eta_{\rm B} 
\equiv n_{\rm baryon}/n_{\rm photon}
=(5.96\pm 0.28)\times 10^{-10}$~\cite{Steigman:2012ve}. 
As demonstrated long ago by Sakharov, the particle physics of the
early universe can explain this asymmetry if B, C, and CP violation
exist in a non-equilibrium environment~\cite{Sakarov:1967dj}. 
Nominally the SM would seem
to possess all the conditions required to generate the BAU. 
However, with 
the discovery of a Higgs boson of 125 GeV in mass, 
the phase transition associated with electroweak symmetry breaking is
no longer of first order~\cite{Aoki:1999fi}, 
and the SM cannot explain a nonzero
BAU. Thus our existence is itself evidence of physics BSM! 
The mechanism of CP violation in the SM has also been faulted,
because an estimate of the BAU (now moot) 
with a sufficiently light
Higgs mass 
yields a BAU orders of magnitude too small, namely, 
$\eta_{\rm B} < 10^{-26}$~\cite{Huet:1994jb}. In the SM nonzero CP-violating 
effects require the participation of 
three generations 
of quarks of differing mass~\cite{Jarlskog:1985ht}. 
Consequently, 
the small value of the computed BAU follows, in part, from 
the smallness of SU(3)$_f$ breaking compared to the electroweak
scale. This special way in which CP 
violation appears in the SM makes it seem that new sources
of CP violation are needed 
to explain the BAU; however, searches for 
such effects at the $B$ factories and through improved EDM limits 
have thus far failed to discover them. 

A BAU could potentially be generated in very different ways, and 
low-energy experiments can 
help select the underlying mechanism.
For example, the discovery of a nonzero EDM at current levels
of sensitivity 
would speak to new CP phases
and the possibility of electroweak baryogenesis. 
The discovery
of neutrinoless $\beta\beta$ decay would tell us that neutrinos are
Majorana particles~\cite{Schechter:1981bd}, 
and would make various models of leptogenesis possible~\cite{Davidson:2008bu}. 
The discovery of $n{\bar n}$ oscillations would reveal that 
 neutrons are also Majorana particles and would support 
alternate models for baryogenesis~\cite{Babu:2013yca}. 
Finally, the discovery of a 
dark-matter asymmetry~\cite{Gardner:2008yn} 
would tell us that DM carries ``baryon'' number, suggesting 
that the key to the nature 
of dark matter and the origin of the BAU 
could be tied~\cite{Davoudiasl:2012uw,Gardner:2013ama}. 
But only EDMs searches are directly connected to the
possibility of new physics at the weak scale. 

\section{Analysis Framework} 
\label{frame}
It is natural to think of the SM as the low-energy limit of a
more fundamental theory, and to use 
an effective theory framework to analyze its possible extensions. 
To illustrate, 
suppose new physics enters at an energy scale $E> \Lambda_{\rm BSM}$. 
Then for energies below the new-physics scale $\Lambda_{\rm BSM}$ 
we can extend the SM through the appearance of effective
operators of mass dimension $D>4$; specifically, 
\begin{equation}
{\cal L} = {\cal L}_{\rm SM} 
+ \sum_i \frac{c_i}{\Lambda_{i}^{D-4}}{\cal O}_i^D\,.
\end{equation}
Noting the severe empirical constraints on new physics 
from flavor-changing 
processes~\cite{Charles:2004jd,Bona:2007vi,Isidori:2010kg}, 
it is efficient to impose 
SU(2)$_L\times$U(1) gauge invariance on the operator basis. 
If we assume that an experimental bound is saturated by a single
term and that the associated $c_i$ is of ${\cal O}(1)$, we can 
estimate the scale $\Lambda_i$; this indicates the rough energy
reach of the experiment. 
For example, a neutrino mass of $0.1\,{\rm  eV}$, 
the expected 
minimum mass accessible to 
near future 
neutrinoless $\beta\beta$ experiments~\cite{deGouvea:2013onf}, if 
generated via the
seesaw mechanism implies 
$\Lambda_{\rm BSM} \sim 10^{14-15}\,{\rm GeV}$~\cite{deGouvea:2013zba}. 
Such estimates 
should be used with care. 

\subsection{The QCD Challenge}
\label{QCD}
Estimates of the energy reach 
of a particular 
experimental measurement can require non-perturbative
QCD input in the form of a hadronic matrix element. 
In this lattice QCD can play a crucial role. 
There are examples, however, where lattice QCD calculations
are not yet good enough to meet experimental needs. 
A prominent example of this 
is the determination of the axial coupling constant of the nucleon  $g_A$. 
In this specific case, $g_A$ can be determined directly from 
experiment, specifically from the measured angular-correlation
coefficients in neutron $\beta$ decay. 
The existing lattice-QCD
calculations do not agree well with each other. Moreover, the lattice 
results typically lie some 5-15\% below the values from $\beta$ decay, 
albeit with much larger errors~\cite{Bhattacharya:2013ehc}. 

\section{Examples}
\label{top}
We now turn to specific examples 
of low-energy  experimental probes of new physics. 

\subsection{Heavy-atom EDMs}
\label{RaRn}
Currently, the most stringent experimental EDM limit comes from the
study of the diamagnetic atom $^{199}$Hg, for which 
$|d| < 3.1\times 10^{-29}\,{\rm e-cm}$ at 95\% C.L.~\cite{Griffith:2009zz}, 
a result roughly a thousand times more sensitive than 
the current experimental limit on the neutron EDM~\cite{Baker:2006ts}. 
However, the atom's electrons shield any nonzero EDM which 
the nucleus may possess and weaken the constraint thereby placed on 
the existence of new sources of CP violation. It has become possible
to study the EDMs of very heavy atoms, such as 
$^{225}$Ra~\cite{Holt:2010upa}
or $^{221/223}$Rn~\cite{Kronfeld:2013uoa}, 
that mitigate the cancelling effect of electron shielding through their
large $Z$, finite nuclear size, and octupole deformation~\cite{Spevak:1996tu}. 
The evasion
of electron shielding in $^{225}$Ra is estimated to be some seven hundred times bigger
than that in $^{199}$Hg~\cite{Dobaczewski:2005hz}, 
making these systems excellent candidates
for the discovery of a nonzero EDM. 
Recently 
the permanent octupole deformation of $^{224}$Ra has been 
established through 
Coulomb excitation studies at REX-ISOLDE (CERN)~\cite{rf:Gaffney2013}; this
makes the nucleus more ``rigid'' and the computation of 
the associated Schiff moment more robust~\cite{Engel:2013lsa}. 
With improved isotope 
yields, as possible, e.g., 
through direct production at a proton linac, 
one expects greatly improved sensitivity to EDMs~\cite{Kronfeld:2013uoa}. 

\subsection{Resolving the limits of the $V-A$ law in $\beta$ decay}
\label{VmA}
The possibility of non-$(V-A)$ interactions in $\beta$ decay
can be probed through the angular correlations of the final-state 
particles. 
Notably the differential decay rate $d^3\Gamma/dE_e d\Omega_{e\nu}$ 
can contain a Fierz interference term $b$; this quantity 
vanishes at tree-level in the SM but is nonzero if scalar or tensor
interactions are present. 
Adopting an effective operator analysis of $\beta$-decay, working
in a SU(2)$_L\times$U(1)-invariant basis in dimension
six~\cite{Buchmuller:1985jz,Grzadkowski:2010es}, 
we have, at the
quark level, 
at low energies~\cite{Cirigliano:2009wk,Bhattacharya:2011qm,Cirigliano:2012ab}, 
\begin{eqnarray}
{\cal L}_{\rm CC}  &=&
- \frac{G_F^{(0)} V_{ud}}{\sqrt{2}} \  \Big[ 
\ \Big( 1   + \delta_{\beta} \Big) \
\bar{e}  \gamma_\mu  (1 - \gamma_5)   \nu_{e}  \cdot \bar{u}   \gamma^\mu  (1 - \gamma_5)  d
\label{eq:leff10}
 \\
&+&  \epsilon_S  \  \bar{e}  (1 - \gamma_5) \nu_{\ell}  \cdot  \bar{u} d
 +
\epsilon_T    \   \bar{e}   \sigma_{\mu \nu} (1 - \gamma_5) \nu_{\ell}    \cdot  \bar{u}   \sigma^{\mu \nu} (1 - \gamma_5) d + \ \dots \ 
+ {\rm h.c.}~. 
\Big]
\nonumber
\end{eqnarray}
The first term represents the famous $V-A$ law of the SM, and the others, 
including the scalar and tensor terms controlled by $\epsilon_S$ and $\epsilon_T$, 
respectively, reflect the appearance of non-SM physics. 
The tree-level coupling $G_F^{(0)}$ is fixed through the measurement
of muon decay and an analysis of its electroweak radiative corrections, 
and $\delta_\beta$ reflect those to semi-leptonic transitions. 
The matching to an effective theory in nucleon degrees of freedom 
requires the computation of hadronic matrix elements; the result maps
to the familiar ${\cal H}_{\rm eff}$ of Lee and Yang~\cite{Lee:1956qn}
employed in Ref.~\cite{Jackson1957zz}.
We refer to  
Ref.~\cite{Cirigliano:2013xha} for a detailed review. 
In neutron $\beta$ decay, we have 
\begin{eqnarray}
&&\!\!\!\!\!\!\!\!\!\!
\langle p(p') | \bar u \gamma^\mu d | n (p) \rangle \equiv
\overline{u}_p (p')
\left[ f_1(q^2)\gamma^\mu - i \frac{f_2(q^2)}{M}\sigma^{\mu\nu}q_\nu +
\frac{f_3(q^2)}{M}q^\mu \right]
u_n(p) \,, 
\nonumber\\
&&\!\!\!\!\!\!\!\!\!\!
\langle p(p') | \bar u \gamma^\mu \gamma_5 d | n (p) \rangle 
\equiv \overline{u}_p (p') \left[
g_1(q^2)\gamma^\mu\gamma_5 - i\frac{g_2(q^2)}{M}\sigma^{\mu\nu}
\gamma_5 q_\nu + \dots
\right]u_n(p) \,, 
\label{eq:form_factorT} \\
&&\!\!\!\!\!\!\!\!\!\!
\langle p(p') | \bar u  d | n (p) \rangle \equiv \overline{u}_p (p') g_S(q^2) u_n(p) \,, \quad
\langle p(p') | \bar u \sigma_{\mu\nu}  d | n (p) \rangle 
\equiv \overline{u}_p (p')
\left[ g_T(q^2)\sigma^{\mu\nu}  + \dots 
\right]u_n(p) \,, 
\nonumber
\end{eqnarray}
where $q\equiv p'-p$ denotes the momentum transfer 
 and $M$ is the neutron mass. 
Working at leading order (LO)
in the recoil expansion 
(i.e., neglecting  terms of ${\cal O}(\varepsilon/M)$, where $|\varepsilon|\ll M$) 
in new physics and at NLO in the SM terms, 
all $q^2$ dependence is negligible --- and other negligible
terms appear as ``$\dots$'' in Eq.~(\ref{eq:form_factorT}). 
Thus we have $f_1(0) \equiv g_V$, $g_1(0) \equiv g_A$ with 
$g_V = 1$ and $f_2(0)=(\kappa_p - \kappa_n)/2$, 
noting $\kappa_{p(n)}$ 
is the anomalous magnetic moment of the proton (neutron), in the SM, 
up to ${\cal O}(\varepsilon/M)$ corrections. 
The quantities
$f_3(0) \equiv f_3$  and $g_2(0) \equiv g_2$ are second-class-current contributions, 
in that they vanish in the SM in the isospin-symmetric limit. 
Bhattacharya et al. have computed $g_S(0)=g_S$ and $g_T(0)=g_T$
in lattice QCD and have shown that their results
sharpen the limits on 
$\epsilon_{S,T}$ considerably~\cite{Bhattacharya:2011qm}. 
Although all the mentioned matrix elements could be computed in lattice QCD, 
not all of the precise matrix elements needed have been --- and we have
already noted the problem with $g_A$. Consequently, to resolve the limits of
the $V-A$ law in $\beta$ decay we must fit for SM physics, specifically 
for $\lambda\equiv g_A/g_V$, and BSM physics
simultaneously~\cite{Gardner:2013aya}. 
There are poorly known recoil-order matrix elements, 
notably $g_2$ and $f_3$; they enter in recoil order and 
can mimic the appearance of scalar and tensor effects. 

\begin{figure}
\centering
\includegraphics[width=7cm,clip]{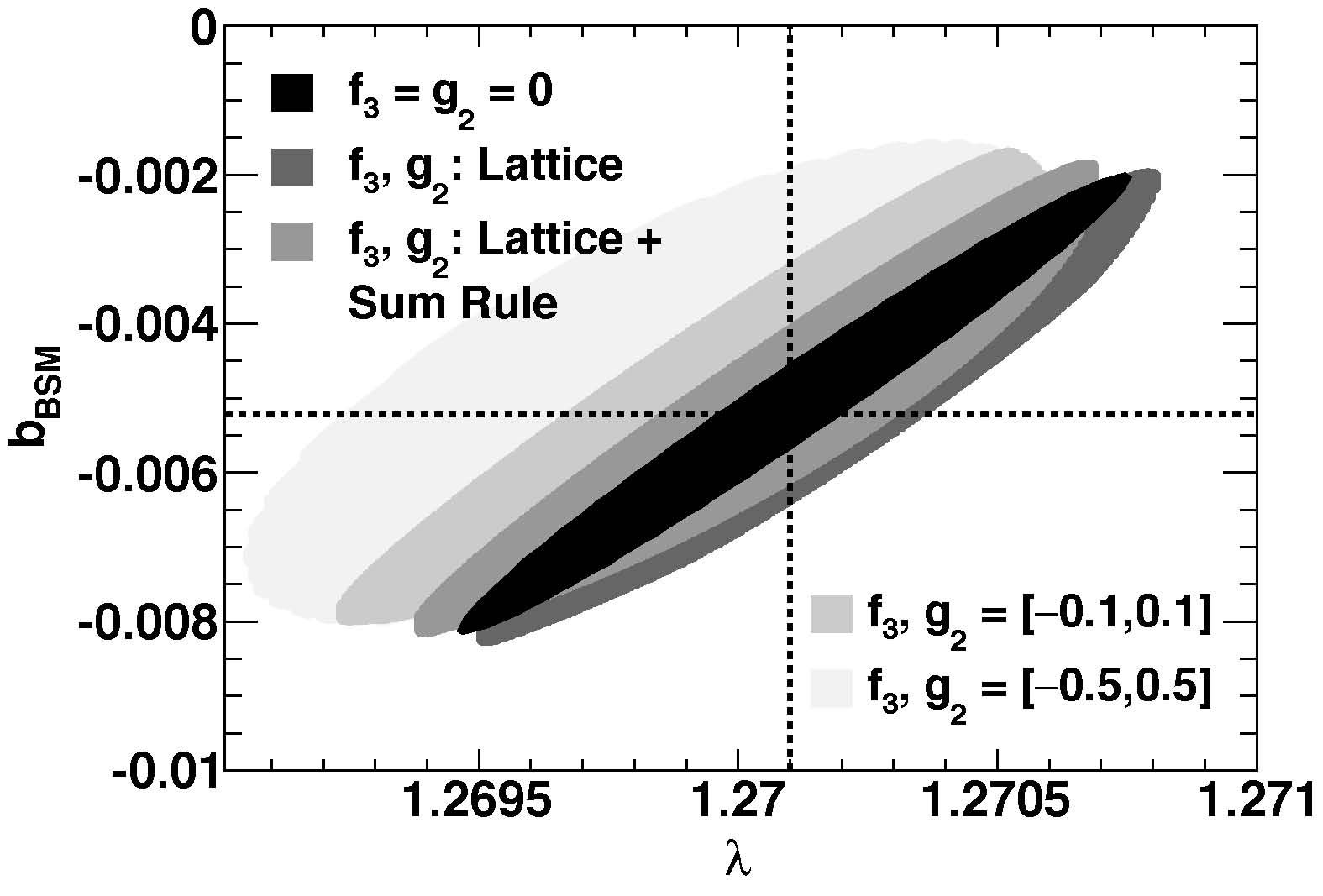}
\includegraphics[width=7.5cm,clip]{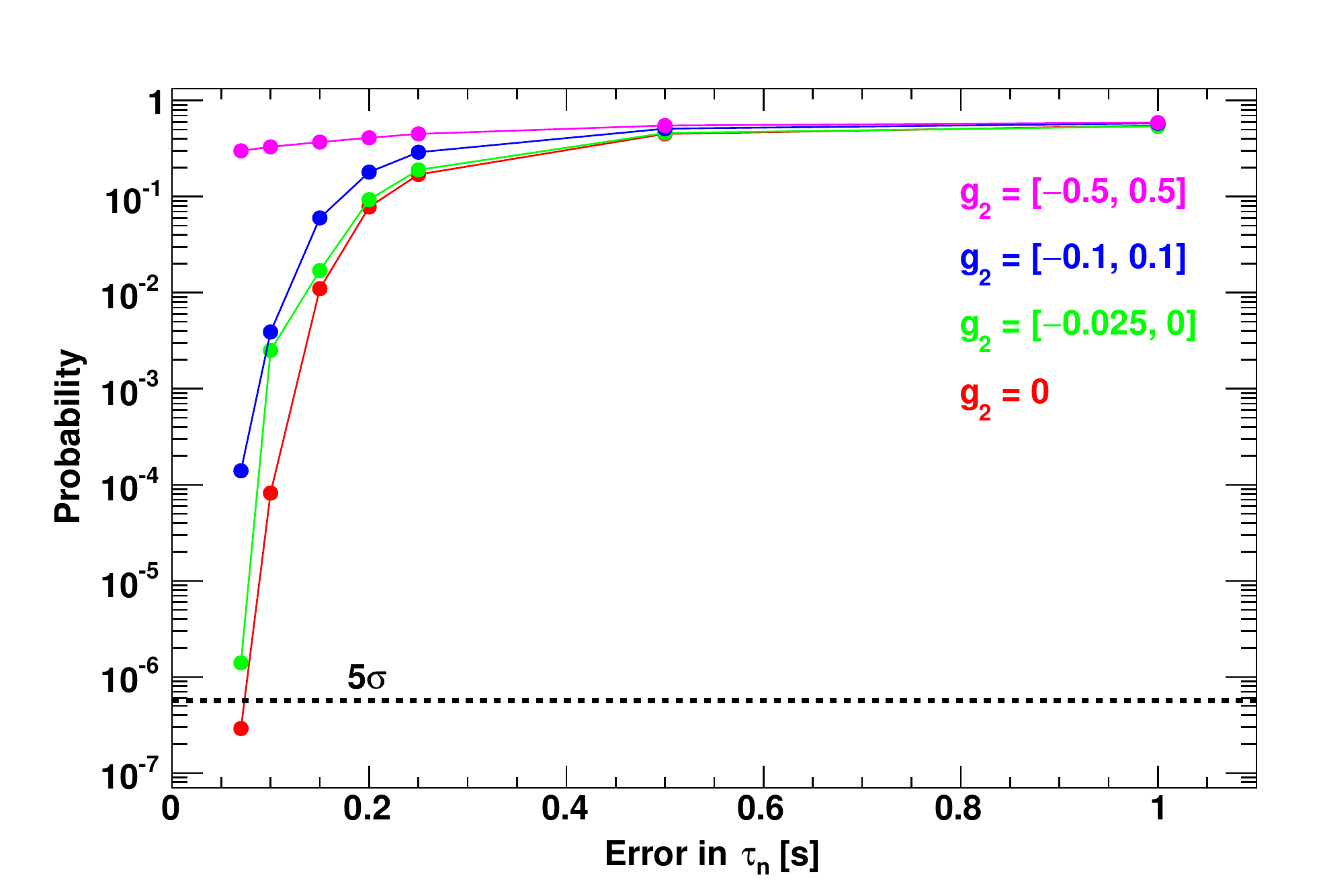}
\sidecaption
\caption{(Left) Illustration of the impact of theoretical certainties
(second-class currents) on the search for 
non-$(V-A)$ currents in neutron $\beta$-decay. We show the results of
the two-parameter $(\lambda,b_{\rm BSM})$ simultaneous fit to the 
$\{ a,A \}$ New Physics data set. The bands indicate the 
68.3\% CL allowed regions. 
(Right) An illustration of the essential role the neutron lifetime 
would play in falsifying 
the SM. We refer to Ref.~\cite{Gardner:2013aya}
for all details. 
}
\label{fig:newphys}       
\end{figure}

Let us consider the prospects for finding BSM physics through $b$; 
we can access this quantity either through a measurement of the electron
energy spectrum or through its impact on the asymmetry measurements
which determine the correlation coefficients $a$ and $A$. 
Many systematic errors
cancel using this latter approach, and we will use it here. 
We address the analysis problem we have posed in the frequentist $R$fit 
(maximum likelihood) framework adopted by CKMFitter for the analysis 
of flavor-changing processes for the parameters of the CKM 
matrix~\cite{Hocker:2001xe,Charles:2004jd}. 
Most importantly this method provides a means
of removing the impact of (SM) theoretical errors on the allowed 
new-physics phase space. We have used Monte Carlo pseudodata of neutron
decay observables to illustrate our implementation of this 
method~\cite{Gardner:2013aya}. For concreteness we recap our methodology. 
We employ a pseudodata set of 
measurements of $a$ and $A$ as
a function of the electron energy $E_e$, along with values of the
neutron lifetime. These results, collectively $\{x_{\rm exp}\}$, 
are to be compared with the
theoretical computations of the same quantities, 
 collectively $\{x_{\rm theo}(y_{\rm mod})\}$, determined by the parameters 
$\{y_{\rm mod}\}$. 
A fraction of the set $\{y_{\rm mod}\}$ can only be determined 
from theory; this subset is labelled $\{y_{\rm calc}\}$. 
The underlying distribution of the $\{y_{\rm calc}\}$ parameters is
ill-known; the test statistic $\chi^2$ is thus modified so that 
the theoretical likelihood does not contribute to the $\chi^2$. 
With this we fit a ``New Physics'' data set for $\lambda$ and 
$b_{\rm BSM}$ in which $\lambda=1.2701$ and 
$b_{\rm BSM}=-0.00522$ for a value of 
$g_T \epsilon_T=1.0\times 10^{-3}$ just below experimental bounds. 
The results as a function of the
theoretical values of $f_3$ and $g_2$ are shown in Fig.~\ref{fig:newphys}. 
We see that  the best-fit ellipses soften in the presence of the
second-class current terms. The method also allows us to construct
a test statistic for the validity of the SM; the essential role
the neutron lifetime plays in realizing it is shown in Fig.~\ref{fig:newphys}. 
For reference, we note that the lattice-QCD result is determined
by 
an extrapolation from the form factors computed in a 
$|\Delta S|=1$ transition~\cite{Sasaki:2008ha}, yielding a result 
at odds with a QCD sum rule calculation~\cite{Shiomi:1996np}. 
For ``Lattice'' we use $f_3 \in (-0.002,0.016)$ and 
 $g_2 \in (0.020,0.066)$, replacing $g_2$ with
$g_2 \in (-0.033,0.066)$ for the union of both. 
We advocate for a lattice QCD calculation of 
$g_2$ and $f_3$ in neutron decay.

\subsection{Spin-independent CP violation in radiative $\beta$ decay}
\label{HHH}
In radiative $\beta$ decay one can form a T-odd correlation from 
momenta alone. This is 
a pseudo-T-odd 
observable, so that it can be mimicked by final-state interactions (FSI)
in the SM. The energy release associated with neutron and
nuclear $\beta$ decay is sufficiently small that only electromagnetic
FSI can possibly generate a mimicking effect. These have been computed 
up to recoil order terms~\cite{Gardner:2012rp}, 
so that we can determine the SM background
rather well. The interaction which generates the primary effect
comes from the gauging of the Wess-Zumino-Witten term under
SM electroweak gauge invariance\cite{Harvey:2007rd,Harvey:2007ca,Hill:2009ek}. 
A direct measurement of 
this correlation constrains the phase of this interaction from 
physics BSM, possibly from ``strong'' hidden sector 
interactions~\cite{Gardner:2013aiw}. 

\section{Summary}
We believe the analysis framework we have espoused
in $\beta$ decay should benefit the analysis of
other low-energy experiments. 
It should be possible to discover physics BSM through
the low energy, precision measurements --- the game is
afoot! 

S.G. would like to thank the organizers for the 
invitation to speak; 
she acknowledges partial support from the U.S. 
Department of Energy Office of Nuclear Physics
under grant number DE-FG02-96ER40989.
B.P. acknowledges partial support from the
U.S. Department of Energy Office of Nuclear Physics
under grant number DE-FG02-08ER41557.

%
\bibliography{svg_menu_biblio}

\begin{thebibliography}{45}

\bibitem{Aad:2012tfa}
G.~Aad et~al. (ATLAS Collaboration), Phys.Lett. \textbf{B716}, 1 (2012),
  \texttt{1207.7214}

\bibitem{Chatrchyan:2012ufa}
S.~Chatrchyan et~al. (CMS Collaboration), Phys.Lett. \textbf{B716}, 30 (2012),
  \texttt{1207.7235}

\bibitem{Ade:2013qta}
P.~Ade et~al. (Planck Collaboration) (2013), \texttt{1303.5081}

\bibitem{Steigman:2012ve}
G.~Steigman, Adv.High Energy Phys. \textbf{2012}, 268321 (2012),
  \texttt{1208.0032}

\bibitem{Sakarov:1967dj}
A.~Sakharov, Pisma Zh.Eksp.Teor.Fiz. \textbf{5}, 32 (1967)

\bibitem{Aoki:1999fi}
Y.~Aoki, F.~Csikor, Z.~Fodor, A.~Ukawa, Phys.Rev. \textbf{D60}, 013001 (1999),
  \texttt{hep-lat/9901021}

\bibitem{Huet:1994jb}
P.~Huet, E.~Sather, Phys.Rev. \textbf{D51}, 379 (1995), \texttt{hep-ph/9404302}

\bibitem{Jarlskog:1985ht}
C.~Jarlskog, Phys.Rev.Lett. \textbf{55}, 1039 (1985)

\bibitem{Schechter:1981bd}
J.~Schechter, J.~Valle, Phys.Rev. \textbf{D25}, 2951 (1982)

\bibitem{Davidson:2008bu}
S.~Davidson, E.~Nardi, Y.~Nir, Phys.Rept. \textbf{466}, 105 (2008),
  \texttt{0802.2962}

\bibitem{Babu:2013yca}
K.~Babu, P.S. Bhupal~Dev, E.C.F.S. Fortes, R.~Mohapatra, Phys.Rev.
  \textbf{D87}, 115019 (2013), \texttt{1303.6918}

\bibitem{Gardner:2008yn}
S.~Gardner, Phys.Rev. \textbf{D79}, 055007 (2009), \texttt{0811.0967}

\bibitem{Davoudiasl:2012uw}
H.~Davoudiasl, R.N. Mohapatra, New J.Phys. \textbf{14}, 095011 (2012),
  \texttt{1203.1247}

\bibitem{Gardner:2013ama}
S.~Gardner, G.~Fuller, Prog.Part.Nucl.Phys. \textbf{71}, 167 (2013),
  \texttt{1303.4758}

\bibitem{Charles:2004jd}
J.~Charles et~al. (CKMfitter Group), Eur.Phys.J. \textbf{C41}, 1 (2005),
  \texttt{hep-ph/0406184}

\bibitem{Bona:2007vi}
M.~Bona et~al. (UTfit Collaboration), JHEP \textbf{0803}, 049 (2008),
  \texttt{0707.0636}

\bibitem{Isidori:2010kg}
G.~Isidori, Y.~Nir, G.~Perez, Ann.Rev.Nucl.Part.Sci. \textbf{60}, 355 (2010),
  \texttt{1002.0900}

\bibitem{deGouvea:2013onf}
A.~de~Gouvea et~al. (Intensity Frontier Neutrino Working Group) (2013),
  \texttt{1310.4340}

\bibitem{deGouvea:2013zba}
A.~de~Gouvea, P.~Vogel, Prog.Part.Nucl.Phys. \textbf{71}, 75 (2013),
  \texttt{1303.4097}

\bibitem{Bhattacharya:2013ehc}
T.~Bhattacharya, S.D. Cohen, R.~Gupta, A.~Joseph, H.W. Lin (2013),
  \texttt{1306.5435}

\bibitem{Griffith:2009zz}
W.~Griffith, M.~Swallows, T.~Loftus, M.~Romalis, B.~Heckel et~al.,
  Phys.Rev.Lett. \textbf{102}, 101601 (2009)

\bibitem{Baker:2006ts}
C.~Baker, D.~Doyle, P.~Geltenbort, K.~Green, M.~van~der Grinten et~al.,
  Phys.Rev.Lett. \textbf{97}, 131801 (2006), \texttt{hep-ex/0602020}

\bibitem{Holt:2010upa}
R.~Holt, I.~Ahmad, K.~Bailey, B.~Graner, J.~Greene et~al., Nucl.Phys.
  \textbf{A844}, 53c (2010)

\bibitem{Kronfeld:2013uoa}
A.S. Kronfeld, R.S. Tschirhart, U.~Al-Binni, W.~Altmannshofer, C.~Ankenbrandt
  et~al. (2013), \texttt{1306.5009}

\bibitem{Spevak:1996tu}
V.~Spevak, N.~Auerbach, V.~Flambaum, Phys.Rev. \textbf{C56}, 1357 (1997),
  \texttt{nucl-th/9612044}

\bibitem{Dobaczewski:2005hz}
J.~Dobaczewski, J.~Engel, Phys.Rev.Lett. \textbf{94}, 232502 (2005),
  \texttt{nucl-th/0503057}

\bibitem{rf:Gaffney2013}
L.P. Gaffney et~al., Nature \textbf{497}, 199 (2013)

\bibitem{Engel:2013lsa}
J.~Engel, M.J. Ramsey-Musolf, U.~van Kolck, Prog.Part.Nucl.Phys.  (2013),
  \texttt{1303.2371}

\bibitem{Buchmuller:1985jz}
W.~Buchmuller, D.~Wyler, Nucl.Phys. \textbf{B268}, 621 (1986)

\bibitem{Grzadkowski:2010es}
B.~Grzadkowski, M.~Iskrzynski, M.~Misiak, J.~Rosiek, JHEP \textbf{1010}, 085
  (2010), \texttt{1008.4884}

\bibitem{Cirigliano:2009wk}
V.~Cirigliano, J.~Jenkins, M.~Gonzalez-Alonso, Nucl.Phys. \textbf{B830}, 95
  (2010), \texttt{0908.1754}

\bibitem{Bhattacharya:2011qm}
T.~Bhattacharya, V.~Cirigliano, S.D. Cohen, A.~Filipuzzi, M.~Gonzalez-Alonso
  et~al., Phys.Rev. \textbf{D85}, 054512 (2012), \texttt{1110.6448}

\bibitem{Cirigliano:2012ab}
V.~Cirigliano, M.~Gonzalez-Alonso, M.L. Graesser (2012), \texttt{1210.4553}

\bibitem{Lee:1956qn}
T.~Lee, C.N. Yang, Phys.Rev. \textbf{104}, 254 (1956)

\bibitem{Jackson1957zz}
J.D. Jackson, S.B. Treiman, H.W. Wyld, Phys. Rev. \textbf{106}, 517 (1957)

\bibitem{Cirigliano:2013xha}
V.~Cirigliano, S.~Gardner, B.~Holstein, Prog.Part.Nucl.Phys. \textbf{71}, 93
  (2013), \texttt{1303.6953}

\bibitem{Gardner:2013aya}
S.~Gardner, B.~Plaster, Phys.Rev. \textbf{C87}, 065504 (2013),
  \texttt{1305.0014}

\bibitem{Hocker:2001xe}
A.~Hocker, H.~Lacker, S.~Laplace, F.~Le~Diberder, Eur.Phys.J. \textbf{C21}, 225
  (2001), \texttt{hep-ph/0104062}

\bibitem{Sasaki:2008ha}
S.~Sasaki, T.~Yamazaki, Phys.Rev. \textbf{D79}, 074508 (2009),
  \texttt{0811.1406}

\bibitem{Shiomi:1996np}
H.~Shiomi, Nucl.Phys. \textbf{A603}, 281 (1996), \texttt{hep-ph/9601329}

\bibitem{Gardner:2012rp}
S.~Gardner, D.~He, Phys.Rev. \textbf{D86}, 016003 (2012), \texttt{1202.5239}

\bibitem{Harvey:2007rd}
J.A. Harvey, C.T. Hill, R.J. Hill, Phys.Rev.Lett. \textbf{99}, 261601 (2007),
  \texttt{0708.1281}

\bibitem{Harvey:2007ca}
J.A. Harvey, C.T. Hill, R.J. Hill, Phys.Rev. \textbf{D77}, 085017 (2008),
  \texttt{0712.1230}

\bibitem{Hill:2009ek}
R.J. Hill, Phys.Rev. \textbf{D81}, 013008 (2010), \texttt{0905.0291}

\bibitem{Gardner:2013aiw}
S.~Gardner, D.~He, Phys.Rev. \textbf{D87}, 116012 (2013), \texttt{1302.1862}

\end{thebibliography}

\end{document}